\documentstyle[12pt]{article}
\newcommand{\be}{\begin{equation}}
\newcommand{\nn}{\nonumber}
\newcommand{\bea}{\begin{eqnarray}}
\newcommand{\eea}{\end{eqnarray}}
\newcommand{\ba}{\begin{array}}
\newcommand{\ea}{\end{array}}
\newcommand{\ee}{\end{equation}}

\expandafter\ifx\csname mathbbm\endcsname\relax

\else

\fi \textheight 22cm \textwidth 15cm \topmargin 1mm \oddsidemargin
5mm \evensidemargin 5mm

\newcommand{\beas}{\begin{eqnarray*}}
\newcommand{\eeas}{\end{eqnarray*}}
\newcommand{\bes}{\begin{equation*}}
\newcommand{\ees}{\end{equation*}}

\newcommand{\lf}{\left}
\newcommand{\ri}{\right}
\newcommand{\f}{\frac}

\def\tr           {\mbox{\rm tr}\,}

\def\i2           {\mbox{$\frac{i}{2}$}}

\def\al           {\alpha}

\def\bet           {\beta}

\def\del           {\delta}

\def\ep           {\epsilon}

\def\ga           {\gamma}

\def\Ga           {\Gamma}

\def\la           {\lambda}

\def\ph           {\phi}
\def\vph           {\varphi}

\def\rh           {\rho}

\def\si           {\sigma}

\def\pl           {\partial}

\begin{document}

\begin{titlepage}
\hfill \vbox{
    \halign{#\hfil         \cr
           } 
      }  
\vspace*{20mm}
\begin{center}
{\LARGE \bf{{On Instantons in Holographic QCD }}}\\ 

\vspace*{15mm} \vspace*{1mm} {Ali Imaanpur}

\vspace*{1cm}

{\it Department of Physics, School of Sciences \\
Tarbiat Modares University, P.O. Box 14155-4838, Tehran, Iran, and\\
Institute for Studies in Theoretical Physics and Mathematics (IPM)\\
P.O. Box 19395-5531, Tehran, Iran \\
Email: aimaanpu@theory.ipm.ac.ir}\\
\vspace*{1mm}

\vspace*{1cm}

\end{center}

\begin{abstract}
We examine instantons and solitons of the effective action of probe D8-branes in the 
background of $N_c$ D4-branes which has served as a holographic description of QCD. 
We show that the 4d instantons sit at the minimum of the Euclidean 5d action. 
Restricting to the static solitons of the five-dimensional model we are led to consider monopoles in a 3-dimensional curved space. Since the background metric depends only on the fifth coordinate, it is possible to reduce the monopole equations to the ones in flat space and write down the explicit solutions. 

\end{abstract}

\end{titlepage}

\section{Introduction}
AdS/CFT correspondence has provided us with a strong/weak duality relating two 
different theories \cite{MAL, W2}. Since its proposal, there has been a great deal of effort to extend the duality to some more realistic theories such as QCD (for instance, see \cite{KAR, ERD, MAT, SUG} and the references therein). Using the duality, one can use the weakly coupled supergravity to learn about the behaviour of the gauge theory at strong coupling. Therefore, in this way, one hopes to better understand some strong coupling phenomena in QCD such as confinement and chiral symmetry breaking.  

In trying to derive an effective action which closely resembles that of QCD, the authors of 
\cite{SUG} have considered a stack of D8-$\overline{{\rm D8}}$ probe branes propagating in the background of $N_c$ D4-branes. Upon compactification over an $S^4$ and setting the 
corresponding components of the gauge fields to zero one is left with an effective 
5-dimensional action. Solitons of this effective theory play a prominent role and are to 
be identified with the baryons in QCD. Static solitons, on the other hand, turn out to be 
the classical solutions of the 4-dimensional Euclidean reduced theory. In \cite{SUG2, YI}, it has been argued that, in a particular limit, the BPST instantons sit at the minima of 
the action. 

In this note we reexamine the 5-dimensional effective action by plugging back 
the metric components to write it in a covariant form. First, by considering only 
those configurations which do not depend on the fifth coordinate $z$, we discuss how 
the 4d instantons can appear as the solutions of the Euclidean 5d action. We perform a coordinate transformation to set $\sqrt{ g_{zz}}=1$ and then show that the flat instantons sit at the minima of the effective 4d action.  
For solitons, we make a different ansatz requiring that fields to be independent of time. 
We write down the 4-dimensional effective action in a covariant form and then look for its minima. Unfortunately, in the reduced 4d action there remains a factor of $\sqrt{-g_{00}}$ which prohibits 4d instantons to solve the field equations. Moreover, this factor cannot be absorbed either by a field redefination or a change of coordinates. There is, however, 
one way out. We observe that if we reduce further the action to 3 dimensions two factors of $\sqrt{-g_{00}}$ and $\sqrt{ g_{zz}}$ cancel each other and one is left with the usual 3d covariant action of gauge fields with no extra factors.  
Upon a field redefinition, we then show that the energy density is minimized 
if the Bogomolny equations are satisfied. 

The organization of this paper is as follows. In section 2, we discuss the instantons in  the Euclidean 5d effective action. By adding a topological term to the action it is seen that 4d instantons sit at the absolute minima of the 5d action. In section 3, we discuss 
the 3-dimensional reduction of the action, and show that the 't Hooft-Polyakov monopole solutions sit at the absolute minima of the action.   
Conclusions are brought in section 4, where we also discuss further directions that can 
be followed.  

\section{Euclidean 5d Action and 4d Instantons}

Before studying the solitons of the model, let us first do a Wick rotation and see 
if we could find instanton like configurations which satisfy the field equations. 
To do so, first we rewrite the Euclidean 5-dimensional effective action of 
D8-branes in a covariant form. We then assume the $z$-independence of the field configurations, and show that the 4d instantons are at the minimum of the 5d action.

Let us start with the effective action of D8-branes in the background of D4-branes, which reads \cite{SUG, SUG2}
\be
S_{\rm {YM}}=-k \int d^4x\, dz\, \tr\! \lf( \f{1}{2}h(z)F_{\mu\nu}^2 + k(z) F_{\mu z}^2 \ri)\, ,
\ee
where
\be
k=\f{\la N_c}{216 \pi^3}\, ,
\ee
and
\be
h(z)=(1 +z^2)^{-1/3}\, ,\ \ \ \ k(z)=1+z^2 \, .
\ee
Notice that $\mu, \nu =1,\ldots, 4$, and $z$ indicates the fifth dimension.
The above action can be rewritten in a covariant form using auxiliary metric components
\be
S_{\rm {YM}}=-k \int \sqrt {g_5}\, d^4x\, dz\,  \tr\! \lf(g^{\mu\rh}g^{\nu\si}F_{\mu\nu}F_{\rh\si} + 
2g^{\mu\nu}g^{zz} F_{\mu z}F_{\nu z} \ri)\, ,\label{FO}
\ee
where the metric components are
\be
g_{\mu\nu}= \f{1}{4} k(z) h(z) \del_{\mu\nu}\, ,\ \ \ g_{zz}= \f{1}{4} h(z)^2 \del_{zz}\, .
\ee
Therefore, the 5-dimensional Euclidean metric is                               
\bea
ds^2 &=& \f{1}{4} k(z) h(z) \del_{\mu\nu}dx^\mu dx^\nu +\f{1}{4} h(z)^2 dz^2\nn \\
&=& \f{1}{4}(1+z^2)^{2/3} \del_{\mu\nu}dx^\mu dx^\nu +\f{1}{4} (1+z^2)^{-2/3} dz^2
\, .\label{5dMET}
\eea
A set of classical solutions of the Euclidean 5d theory can be recognized as follows.  
The classical action (\ref{FO}) shows that if we set $F_{\mu z}=0$, and require that 
all other fields to be independent of $z$ then (\ref{FO}) is reduced to a 4d action 
with instantons as its classical minima. However, we notice that the reduced 4d action  
will have an extra $\sqrt {g_{zz}}$ factor compared to the Yang-Mills action in curved 
background. To get rid of that factor, and before reduction to four dimensions, we do 
a coordinate transformation, $z\to z'$, and write the 5d metric (\ref{5dMET}) as follows
\be
ds^2 = f(z')\lf( \del_{\mu\nu}dx^\mu dx^\nu\ri) + dz'^2 
\, ,
\ee
where 
\be
z'(z) =\f{1}{{2}}\int \f{dz}{(1+z^2)^{1/3}}= \f{1}{{2}}
\, z\, F\lf(\f{1}{2}, \f{1}{3}; \f{3}{2}; -z^2\ri) \, ,
\ee
with $F$ the hypergeometric function, and
\be
f(z')= \f{1}{4}(1+z(z')^2)^{2/3}\, .\label{INV2}
\ee 
Therefore, in this coordinate system $\sqrt{ g_{z'z'}}=1$, and the action reads 
\be
S_{\rm {YM}}=-k \int \sqrt {g_5}\, d^4x\, dz\,  \tr\! \lf(g^{\mu\rh}g^{\nu\si}F_{\mu\nu}F_{\rh\si} \ri)
\, .
\ee
The absolute minima of the 4d action can be worked out by adding and subtracting a topological term proportional to the instanton number:
\bea
S&=& -k \int \sqrt{g_4}\, d^4x \,  \tr\! \lf(g^{\mu\rh}g^{\nu\si}F_{\mu\nu}
F_{\rh\si} \ri)\nn \\
&=& -\f{k}{2} \int \, \sqrt{g_4}\, d^4x \,  \tr\! \lf(F_{\mu\nu} - \f{1}{2\sqrt{g_4}}g_{\mu\rho} g_{\nu\si}\ep^{\rho\si\lambda\del}{F}_{\lambda\del} \ri)^2\nn \\
\ \ \ \ \ \ &-&\f{k}{2} \int \, d^4x \,  \tr\! \lf(\ep^{\rho\si\lambda\del}
F_{\rho\si}{F}_{\lambda\del}\ri) \, .
\eea
Written in this form, it is now clear that the absolute minima of the 4d action are the 
instantons on a curved 4-dimensional space with the metric $g_{\mu\nu}$:
\be
{F}_{\mu\nu}=\f{1}{2\sqrt{g_4}}g_{\mu\rho} g_{\nu\si}\, \ep^{\rho\si\lambda\del}{F}_{\lambda\del} \, . 
\ee
with the convention $\ep^{123z}=1$ and $\ep_{123z} \sim g_4$. However, since we are in 
four dimensions and since the metric in the remaining 4 coordinates 
is conformally flat, the above equations reduce to the instanton equations on flat space
\be
{F}_{\mu\nu}=\f{1}{2}\del_{\mu\rho}\del_{\nu\si}\, \ep^{\rho\si\lambda\del}
{F}_{\lambda\del} \, ,
\ee
with completely known solutions.

\section{String-like Solutions and Monopoles} 
In this section, we discuss a class of solutions of the 5d theory which are independent of $t$ and the $z$ direction, and hence resemble vortex solutions. The main observation 
is that these strnig solitons can be identified with 3d monopoles in the transverse directions. In what follows, first we reduce 
the action to a covariant Euclidean 4d action, and show that 4d instantons cannot 
solve the field equations. Then, we further reduce the action to 3 dimensions and 
notice that, because of the form of the metric, the monopole solutions are in fact 
at the minima of the 3d action.  

The 5d action that we start with is different from that of previous section in that it includes adjoint scalars $\ph$. In \cite{SUG, SUG2}, these have been set to zero and the $U(1)$ part of $\ph$ gets stabilized at antipodal points of $S^1$ (which is orthogonal to the worldvolume of D8-branes). So including these, the 5d action is  
\be
S_{\rm {YM}}=-k \int \sqrt {g_5}\, d^5x \,  \tr\! \lf(g^{ML}g^{NK}F_{MN}F_{LK} + 2 
g^{MN}D_{M}\ph D_{N}\ph \ri)\, , \label{AC5}
\ee
In looking for solitons of the 5d model, the simplest choice is to look for a static field configuration and set $\pl_0 =A_0 =0$,  so the action reads
\be
S_{\rm {YM}}=-k \int \sqrt {-g_{00}}\, dt\, \sqrt {g_4}\, d^3x dz\,  \tr\! \lf(g^{\al\ga}g^{\bet\del}F_{\al\bet}F_{\ga\del} + 2 
g^{\al\bet}D_{\al}\ph D_{\bet}\ph \ri)\, , \label{AC4}
\ee
with $\al ,\bet, \ldots =1,2,3,z$, and the four-dimensional metric                             
\bea
ds^2 =\f{1}{4}(1+z^2)^{2/3} \del_{ij}dx^idx^j +\f{1}{4} (1+z^2)^{-2/3} dz^2
\, .\label{MET}
\eea

The solitonic solutions now have to minimize the action in (\ref{AC4}). As in the previous 
section, if $\sqrt {-g_{00}}$ was absent in (\ref{AC4}), we could have argued that the 
instantons (with $\ph =0$) in the above metric background are sitting at the minima of the action. However, in this case it is not possible to get rid of $\sqrt {-g_{00}}$ by a coordinate transformation, and thus instantons are not solutions to the field equations. This is 
easily seen as follows.\footnote{ We thank A. Karch for pointing out an error in field 
equations in the first version of this paper.} 
By varying the action (\ref{AC4}), the field equations read 
\be
{\hat D}_\al \lf(\sqrt {-g_{00}}\, F^{\al\bet}\ri) +i\sqrt {-g_{00}}
\, \lf[\ph\, ,{\hat D}^\bet\ph \ri]=0\, , \label{FIELD}
\ee
\be
{\hat D}_\al \lf(\sqrt {-g_{00}}\, {\hat D}^\al\ph \ri)=0\, .
\ee
with ${\hat D}_\al$ the 4d connection. Now for instantons we have
\be
F_{\al\bet}=\f{1}{2\sqrt {g_4}}g_{\al\ga}g_{\bet\eta}\ep^{\ga\eta\del\kappa}F_{\del\kappa} \, ,\label{INST}
\ee
together with $\ph =0$. These clearly do not satisfy (\ref{FIELD}); with the $\sqrt {-g_{00}}$ factor inside the covariant derivative the field equations  do not reduce to the Bianchi identities. 

Although $\sqrt {-g_{00}}$ cannot be set to $1$ by a coordinate transformations, we can get 
rid of that by dimensionally reducing the action one step further. In fact, since
\be
\sqrt {-g_{00}}\cdot \sqrt {g_{zz}}=\f{1}{4} \, ,\label{gg}
\ee 
we observe that if we reduce action (\ref{AC4}) to three dimensions there is a chance of reducing the field equations to some first order differential equations. Let us first  
discuss that without the scalars $\ph$ it is not possible to get to the monopole equations. 

\subsection{ The case with $\ph =0$}
 
Let us first set $\ph=0$ and assume $\pl_z A_i=0$. We can absorb $\sqrt {g^{zz}}$ into  $F_{iz}$ and write 
\be
\sqrt {g^{zz}}F_{iz}= 2(1+z^2)^{1/3}D_i A_z 
\equiv  D_i\, {\vph} \, , \label{ANSATZ}
\ee
where we have defined
\be
{\vph} =2(1+z^2)^{1/3}A_z \, .
\ee
Action (\ref{AC4}) now becomes
\be
S_{\rm {YM}}= -\f{k}{4} \int dz\, \, \sqrt {g_3}\, d^3x \,  \tr\! \lf(g^{im}g^{jn}F_{ij}F_{mn} + 
2g^{ij} D_{i}{\vph} D_{j}{\vph} \ri)  \, , \label{E}
\ee
where $i,j,\ldots =1,2,3$. Notice that if we vary the above action, in contrast to what we saw in (\ref{FIELD}), we will obtain the ordinary covariant 3d field equations without any extra factor. It is 
amusing to see whether the solutions to the 3d equations of the above action are in fact solutions of the 5d action in (\ref{FO}). Let $\tilde {D}_M$ and $\hat {D}_\al$ indicate the five and four dimensional connections respectively, $M$ runs from 
$0,\ldots ,4$ and $\al$ from $1,\ldots ,4$. For a static solution of 5-dimensional theory 
we require $A_0=0$ and set $F_{\al 0}=0$, then the 5d equations of motion 
split
\be
\tilde {D}_\al F^{\al 0}=0\, ,\ \ \ \ \tilde{D}_0 F^{0\bet} + 
\tilde{D}_\al F^{\al\bet}=0\, .
\ee
Since $F_{\al 0}=0$, the first equation is satisfied and the second one reduces to the 4d equation of motion of the action in (\ref{AC4}) (with $\ph =0$):
\be
\tilde{D}_\al F^{\al\bet} + \Ga^0_{0z} F^{z\bet}=
\hat {D}_\al F^{\al\bet}+ \Ga^0_{0z} F^{z\bet}=0\, .\label{DD}
\ee
Now if we set $\bet =z$, since $F_{\al 0}=0$ and $\Ga^z_{zi}=0$, we have
\be
\hat {D}_\al F^{\al z}= {D}_i F^{iz}=0 \ \ \Rightarrow \ \ D_iD^i\, \vph =0\, .
\ee
which is the $\ph$ equation of motion derived from the 3d action (\ref{E}). 
However, if we set $\bet =i$ in (\ref{DD}) we will have:
\be
\hat {D}_\al F^{\al i}+ \Ga^0_{0z} F^{z i} ={D}_j 
F^{j i}+ \pl_zF^{zi} +\Ga^j_{jz} F^{zi} + i[A_z , F^{zi}]=0 \, ,\label{ANS}
\ee
where in deriving the second equality we used the fact that $\Ga^0_{0z}= -\Ga^z_{zz}$, 
and $\Ga^z_{zi}=0$. For this to be the equation of motion for $A_i$ derived from the action 
(\ref{E}) we must have
\be
\pl_zF^{zi} +\Ga^j_{jz} F^{zi}=0\, ,
\ee
which in turn implies
\be
F_{zi} \sim \f{1}{(1+z^2)}\, .
\ee
Further, recall that $\pl_z A_i=0$, so 
\be
A_{z} \sim \f{1}{(1+z^2)}\ \ \Rightarrow \ \ [A_z , F^{zi}] \sim \f{1}{(1+z^2)^2}\, ,
\ee
On the other hand, since $F^{ij} \sim (1+z^2)^{-4/3}$, we see that the above $z$-dependence is not consistent with the equation of motion in (\ref{ANS}). Hence we conclude that our 
ansatz (\ref{ANSATZ}) is not consistent with the 3d equations of motion. The only choice 
is to set $F_{zi}=0$, for which (\ref{ANS}) reduces to the ordinary pure Yang-Mills equations in 3 dimensions
\be
D_j F^{ji} =0\, .
\ee  

\subsection {The case with $\ph \neq 0$}

The equations of motion from (\ref{AC5}) read
\be
{\tilde D}_M F^{MN} + i[ \ph\, , D^N\ph]=0\, ,
\ee
\be
D_MD^M \ph =0\, .
\ee
requiring $\pl_0 =A_0 =0$, the above equations to four dimensions
\be
{\hat D}_\al F^{\al\bet}+ \Ga^0_{z0}F^{z\bet} + i[ \ph\, , D^\bet\ph]=0\, ,
\label{EQ4}
\ee
\be
D_\al D^\al \ph =0\, .
\ee
Splitting the indices to $z$ and $i$ the equations (\ref{EQ4}) read
\be
D_i F^{iz} +i[\ph\, , D^z\ph]=0\, ,\label{E11}
\ee
\be 
{D}_j F^{ji}+ i[A_z, F^{zi}] + \pl_zF^{zi} +\Ga^j_{jz}F^{z i} + i[ \ph\, , D^i \ph]=0
\label{E22}\, .
\ee
in deriving the second equality we used the fact that $\Ga^0_{0z}= -\Ga^z_{zz}$.
We now further reduce these equations to three dimensions as follows. We notice that there 
is a consistent ansatz of the form $\pl_zA_i=0$ and $A_z=0$, which also implies $F_{zi}=0$. 
In this case equations (\ref{E11}) and (\ref{E22}) reduce to
\be
[\ph\, , \pl^z\ph]=0\, ,\label{E12}
\ee
\be 
{D}_j F^{ji} + i[ \ph\, , D^i \ph]=0\, .
\ee
In three dimensions, the last equation is solved by monopole solutions which satisfy the 
first order Bogomolny (monopole) equations:
\be
F_{ij}=\f{1}{\sqrt {g_3}}g_{im}g_{jn}\ep^{mnk}D_{k}{ \ph}\, ,\label{BOG}
\ee
plugging back the metric components this equation becomes
\be
F_{ij}=\f{1}{2}(1+z^2)^{1/3}\del_{im}\del_{jn}\ep^{mnk}D_{k}{ \ph}\, .\label{MM}
\ee
Since we required $\pl_zA_i=0$, the left hand side is $z$-independent. So we are lead to 
define 
\be
{\tilde \ph} = \f{1}{2}\, (1+z^2)^{1/3}{ \ph}\, ,
\ee
which is to be $z$-independent. Note that by this, equation (\ref{E12}) is trivially 
satisfied. Written in terms of $\tilde \ph$, equation (\ref{MM}) turns to the monopole 
equation on flat 3d space:    
\be
F_{ij}= \del_{im}\del_{jn}\ep^{mnk}D_{k}{\tilde \ph}\, .\label{MMM}
\ee

These monopole configurations can be seen that minimize the energy 
density (or tension of the string). Recalling (\ref{gg}), and with our ansatz for the 
time and $z$ independence of $A_i$ and ${\tilde{ \ph}}$, the reduced 3d action can be 
read from (\ref{AC5}): 
\be
S_{\rm {YM}}=-k \int \sqrt {g_3}\, dt\, d^3x\, dz \,  \tr\! \lf(g^{im}g^{jn}F_{ij}F_{mn} + 2 
g^{ij}D_{i}\ph D_{j}\ph \ri)\, .
\ee
Hence, for the energy density we have 
\bea
E \!\!\! &=&\!\!\! \f{k}{4} \int \! \sqrt {g_3}\, d^3x \,  \tr\! \lf(g^{im}g^{jn}F_{ij}F_{mn} + 
2 g^{ij} D_{i}{\ph} D_{j}{\ph} \ri) \nn \\
&=&\!\!\! \f{k}{4} \int \! \sqrt {g_3}\, d^3x \,  \tr\! \lf(F_{ij}-\f{1}{\sqrt {g_3}}g_{im}g_{jn}\ep^{mnk}D_{k}{\ph} \ri)^2 \! 
+\! \f{k}{2} \int \! d^3x \,  \tr\!\! \lf(\ep^{ijk}F_{ij}D_{k}{\ph} \ri)\label{ee} 
\eea
The last term is proportional to the winding number; a topological number which is the 
same for all field configurations having the same boundary conditions. Therefore, the energy density functional, in each topological sector, is minimized if the fields satisfy the Bogomolny (monopole) equations (\ref{BOG}). For such configurations, the energy density (energy per unit invariant length in the $z$ direction) is  
\be
\f{E}{\sqrt{g_{zz}}} =  
2k \int \, d^3x \,  \tr\!\! \lf(\ep^{ijk}F_{ij}D_{k}{\tilde {\ph}} \ri)\, , 
\ee
which is finite and proportional to the winding number associated to the behaviour of 
$\tilde {\ph}$ on the boundary of $R^3$. Therefore, these monopole solutions, viewed from the four dimensions, look like strings extended along the $z$ direction.  

The general solutions to the monopole equations (\ref{MMM}) can be found through the 
Nahm's construction \cite{NAM, NEK}. For reference purposes, here we write down the 
explicit solutions for the monopoles of charge one and with the $SU(2)$ gauge group. 
For the scalar and gauge fields they read
\[
\tilde {\ph} = \f{1}{2}\lf( \f{a}{\tanh (ra)} -\f{1}{r}\ri) \si_3\, ,
\]
\[
A_+ = \lf(
\begin{array}{clrr} %
\f{ 1 -x_3/r}{4x_+}& \f{ax_-(x_+x_- + 2x_3(x_3 - r)) {\rm Csch} (ar)}
{4r(r-x_3)^{3/2}(r+x_3)^{1/2}} \\       
 \f{ax_-(x_+x_- + 2x_3(x_3 + r)){\rm Csch}(ar)}{4r(r-x_3)^{1/2}(r+x_3)^{3/2}} & 
\ \ \ \  \f{ 1 +x_3/r}{4x_+}
\end{array}\ri)\, ,
\]
\[
A_- = \lf(
\begin{array}{clrr} %
\f{ 1 -x_3/r}{4x_-}&        
 \f{ax_+(x_+x_- + 2x_3(x_3 + r)){\rm Csch}(ar)}{4r(r-x_3)^{1/2}(r+x_3)^{3/2}}\\
 \f{ax_+(x_+x_- + 2x_3(x_3 - r)) {\rm Csch} (ar)}
{4r(r-x_3)^{3/2}(r+x_3)^{1/2}} & 
\ \ \ \  \f{ 1 +x_3/r}{4x_-}
\end{array}\ri)\, ,
\]
\[
A_3 = \f{a \sqrt{r^2-x_3^2}}{2r\sinh (ar)}\lf(
\begin{array}{clrr} %
0 &  1      
 \\
-1  & 0
\end{array}\ri)\, ,
\]
with $x_{\pm}=x_1\pm i x_2$, $r=\sqrt{\sum_{i=1}^3x_i^2}$, and $a$ an arbitrary constant 
specifying the boundary value of $\ph$. The field strength components are  
computed to be
\[
F_{12}= \f{1}{2r}\lf\{\lf( \f{1}{r^2}-\f{a^2}{\sinh^2(ar)}\ri)x_3\si_3 
-ia\f{\sqrt{r^2 -x_3^2}}{\sinh (ar)}\lf( \f{a}{\tanh (ra)} -\f{1}{r}\ri) \si_1\ri\} ,
\]
\[
F_{23}= \f{1}{2r}\lf\{\lf( \f{1}{r^2}-\f{a^2}{\sinh^2(ar)}\ri)x_1\si_3 
+\f{ia(x_1x_3 \si_1 - rx_2\si_2)}{\sqrt{r^2 -x_3^2}\sinh (ar)}\lf( \f{a}{\tanh (ra)} -\f{1}{r}\ri)\ri\} , 
\]
\[
F_{31}= \f{1}{2r}\lf\{\lf( \f{1}{r^2}-\f{a^2}{\sinh^2(ar)}\ri)x_2\si_3 
+\f{a(x_2x_3 \si_1 + rx_1\si_2)}{\sqrt{r^2 -x_3^2}\sinh (ar)}\lf( \f{a}{\tanh (ra)} -\f{1}{r}\ri)
\ri\} .
\]

\section{Conclusions and Outlook}

In this paper, we examined the 5-dimensional effective action of D8-branes in the 
background of D4-branes. We made two main observations. In the case of Euclidean 
5d theory, by a coordinate transformation we showed that the 4d flat instantons minimize 
the action. For static solitons, we derived the effective 4-dimensional 
action and argued that instantons are not minimizing the action. On the 
other hand, we showed that upon a further reduction to 3 dimensions monopoles 
appear as absolute minima of the energy density. The form of the metric allowed us 
to convert the monopole equations to the ones on flat space so that we were able to write down the explicit solutions.

As for comparison with the baryons in QCD, it is interesting to study the quantization of the collective modes. The contribution of the CS term is another subject that is necessary to be discussed along the lines of \cite{SUG2}. One interesting aspect of our solutions, 
in contrast to the solutions in \cite{SUG2}, is that they satisfy the equations of motion 
without having to take a particular limit on the moduli parameters like $a$.

\vspace{5mm}
     
\hspace{40mm}

\hspace{-6mm}{\large \textbf{Acknowledgment}}

\vspace{1.5mm}

\noindent
This work was initiated by a set of interesting lectures on holographic QCD given by 
A. Dhar in IPM String School and Workshop ISS2007, Tehran, Iran.

\end{document}